\documentclass[onecolumn,prl,superscriptaddress,showpacs,byrevtex]{revtex4}

\usepackage{graphicx,amssymb,amsmath}
\usepackage{natbib}

\begin{document}
\title{Excitability in the vicinity of a saddle-node bifurcation: a mechanism for  reversals.}

\author{Fran\c{c}ois PETRELIS and St\'ephan FAUVE}
\affiliation{Laboratoire de Physique Statistique, CNRS UMR 8550, Ecole Normale Sup\'erieure,\\ 24 rue Lhomond, 75005 Paris, France}

\label{firstpage}

\begin{abstract}
We present a new mechanism for oscillatory or random reversals of the magnetic field that occur from a stationary dynamo state. The basic ingredient is the existence of two nearly critical dynamo eigenmodes, a stable and an unstable one, that collide during a saddle-node bifurcation. Above this bifurcation, a finite amplitude limit cycle is generated. In the neighbourhood of the bifurcation, even a small amount of turbulent fluctuations is enough to generate random reversals of the magnetic field. When the two competing eigenmodes have different symmetries, this scenario requires a broken symmetry of the flow that generates the dynamo. This is in good agreement with the observations of the VKS experiment. It can also explain the dynamics of planetary or stellar magnetic fields and can be used to understand reversals of large scale flows that often develop on a turbulent background.
\end{abstract}

\maketitle

\section{Introduction}

Palaeomagnetic measurements show that the Earth magnetic field has a fixed polarity for long durations, but from time to time, it flips with the poles reversing sign
(see the reviews by Dormy \textit{et al.} (2000), Roberts \& Galtzmaier (2000)). 
The first simple models of field reversals involved disk dynamos (Rikitake 1958) or drastic truncations of the magnetohydrodynamic (MHD) modes (Nozieres 1978). A different class of models, also involving a few coupled differential equations, is based on the assumption that several magnetic eigenmodes are competing above the dynamo threshold (Tobias \textit{et al.} 1995, Knobloch \& Landsberg 1996). In this framework, it has been proposed to relate reversals to trajectories close to heteroclinic cycles that connect unstable fixed points $\pm \bf {B}$ (Armbruster \textit{et al.} 2001, Melbourne  \textit{et al.} 2001). 
Other models take into account turbulence in a stronger way by modelling it through random fluctuations of one of the coefficients of the dynamical system (Hoyng \& Duistermaat 2004). A different approach, initiated by Parker (1969), consists in trying to identify properties of the velocity field that generate dynamics of the magnetic field. Recent numerical simulations have modelled the effect of the velocity with a fluctuating  $\alpha$-effect (Giesecke  \textit{et al.} 2005, Stefani \& Gerbeth 2005). 
Since 1995, three dimensional numerical simulations of the MHD equations in a rotating sphere have been able to simulate a magnetic field that displays reversals (see the reviews by Dormy \textit{et al.} 2000, Roberts \& Galtzmaier 2000, or for more recent works,  Li \textit{et al.} 2002, Kutzner \& Christensen 2002, Wicht \& Olson 2004). The experimental observation of reversals of the magnetic field and of other time dependent regimes has been performed only recently in a turbulent swirling flow of liquid sodium  (VKS experiment) (Berhanu \textit{et al.} 2007, Ravelet \textit{et al.} 2008).

The purpose of this paper is to give a new low dimensional description of the evolution of the magnetic field generated by dynamo action. From this description, different dynamical regimes of the magnetic field can be identified and their properties can be predicted. These predictions are in good agreement with the VKS experiment. They can be also tested against the results observed in numerical, geophysical and astrophysical dynamos. 

\section{Modes and symmetries}

\subsection{Description of the VKS experiment.}

\begin{figure}[!htb] 
\centerline{\includegraphics[width=1\textwidth]{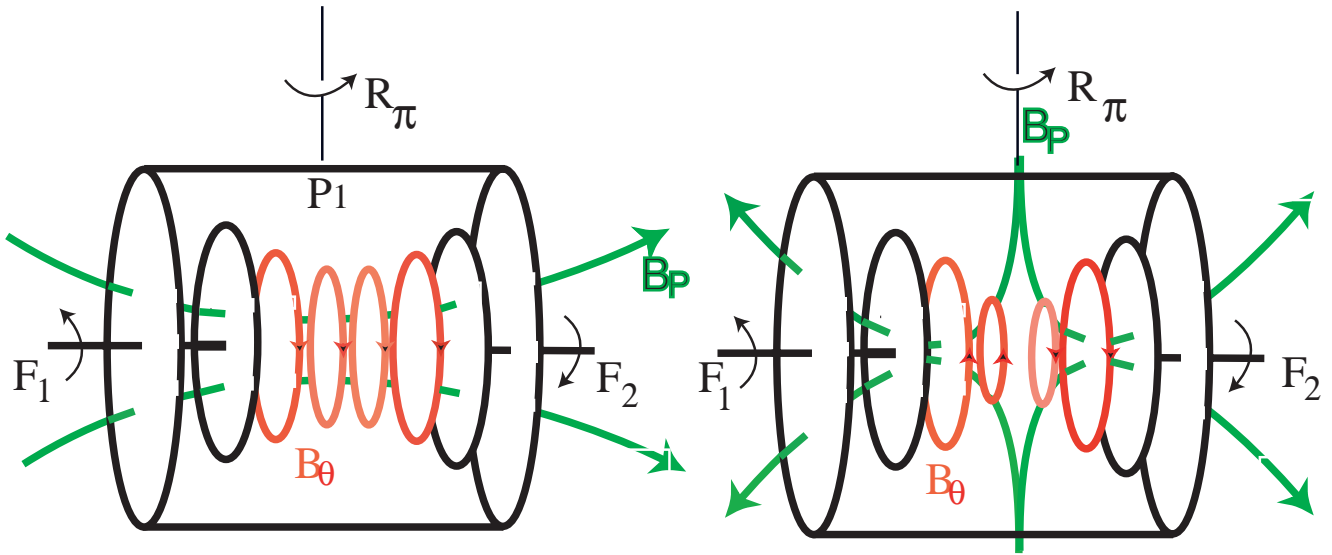}}
\vspace{-2 cm}
\caption {Possible eigenmodes  of the VKS experiment. The two disks counter-rotate with frequency $F_1$ and $F_2$. Left: magnetic dipolar mode. Right: magnetic quadrupolar mode. Poloidal (green) and toroidal (red) components are sketched.}
\label{figmodevks}
\end{figure}

The VKS experiment involves a turbulent swirling flow of liquid sodium, generated by  two impellers, counter-rotating at frequency $F_1$ (respectively $F_2$) in an inner copper cylinder, as sketched in fig. \ref{figmodevks} (see Monchaux \textit{et al.} 2007 for the experimental set-up). 
When the disks counter-rotate with the same frequency $F$, a statistically stationary magnetic field is generated when $F$ is large enough. Its mean value involves a dominant poloidal dipolar component, ${\bf B}_P$, along the axis of rotation, together with a related azimuthal component $\bf{B}_{\theta}$, as displayed in fig. \ref{figmodevks} (left). This shows that the VKS dynamo is not generated by the mean flow alone that would give a non axisymmetric magnetic field according to Cowling theorem. As explained in P\'etr\'elis  \textit{et al.} (2007), a possible mechanism is of $\alpha$-$\omega$ type, the $\alpha$-effect being related to the helical motion of the radially expelled fluid between two successive blades of the impellers, and the $\omega$-effect resulting from differential rotation due to counter-rotation of the impellers. 

A striking  observation of the VKS experiment is that time dependent magnetic fields are generated only when the impellers rotate at different speeds (Berhanu \textit{et al.} 2007).  
When the rotation frequency difference $f = F_1-F_2$ is increased, the dynamo first remains stationary but the relative amplitudes of the field components are modified. The field is measured at position $P1$ in fig. \ref{figmodevks}. A dipolar (respectively quadrupolar) field is associated to a field along $B_x$ (respectively $B_r$). Increasing  the velocity difference from zero results in a continuous modification of the magnetic field from dipole to quadrupole. For a larger value of $f$, a bifurcation takes place and  the field  becomes  time-periodic. The range of existence of this regime is small and if $f$ is further increased, a stationary magnetic field is recovered. Remarkably, the stationary solutions are associated to points in the phase space that are located on the limit cycle related to the oscillatory solution. These bifurcations have been described in detail (Ravelet \textit{et al.} 2008). It has been also shown that the parameter space $(F_1, F_2)$ consists of several regions with an alternation of time dependent and stationary dynamos. Finally, a direct bifurcation from a non-dynamo state to a time dependent dynamo has been never observed. 

\subsection{An amplitude equation for dipolar and  quadrupolar modes}

As said above, the most striking  feature of the VKS experiment is that time dependent magnetic fields are generated only when the impellers rotate at different speeds. We will show that this can be related to an additional symmetry. When $F_1 = F_2$,
the experiment is invariant with respect to a rotation of $\pi$ around any axis located in the midplane between the two disks. Let ${\cal R}_{\pi}$ be one of these rotations.  We expect that in the counter-rotating regime, the modes involved in the dynamics are either symmetric or antisymmetric. Such modes are displayed in figure  \ref{figmodevks}. A dipole is changed to its opposite by ${\cal R}_{\pi}$, whereas a quadrupole is unchanged. More generally, we name ``dipole" (respectively ``quadrupole"), modes with dipolar (respectively quadrupolar) symmetry even though they might involve a  more complex spatial structure.

We assume that the magnetic field is the sum of a dipolar component with an amplitude $D$ and a quadrupolar one, $Q$. We define $A=D+i\, Q$ and we assume that an expansion in power of  $A$  and its complex conjugate $\bar{A}$ is pertinent close to threshold in order to obtain an evolution equation for both modes. Taking into account the invariance ${\bf B} \rightarrow-{\bf B}$, {\it i. e.} $A\rightarrow -A$,  we obtain 
\begin{eqnarray}
\dot{A}&=&\mu A+\nu \bar{A}+\beta_1 A^3+\beta_2 A^2\bar{A}+\beta_3 A \bar{A}^2+\beta_4\bar{A}^3\,,
\label{eqdipquad}
\end{eqnarray}
where we limit the expansion to the lowest order nonlinearities.
In the general case, the coefficients are complex and depend on the experimental parameters.

Symmetry of the  experiment  with respect to ${\cal R}_{\pi}$ when the disks exactly counter-rotate, amounts to constraints on the coefficients. Applying this transformation to the magnetic modes changes $D$ into $-D$ and  $Q$ into $Q$, thus $A\rightarrow-\bar{A}$. We conclude that, in  the case of exact counter-rotation, all the coefficients are real. When the frequency difference $f$ is increased from zero, assuming that the coefficients are analytical functions of $f$, we obtain that the real parts of the coefficient are even and the imaginary parts are odd functions of $f$.

When the coefficients are real, the growth rate of the dipolar component is $\mu_r+\nu_r$ and that of the quadrupolar component is $\mu_r-\nu_r$. The dipole being obtained for exact counterrotation implies that $\nu_r > 0$ for $f=0$. By increasing $f$, we expect that $\nu_r$ changes sign and favors the quadrupolar mode.

\subsection{Study of the amplitude equation}

To analyse the properties of eq. \ref{eqdipquad}, we write  the equations for the  phase $\theta$ and the modulus $r$  of  $A$, $A=r\,\exp{(i\,\theta)}$, 
\begin{eqnarray}
\dot{r}&=&r\large(\mu_r+\beta_{2r}\, r^2+\cos{(2\theta)}\,(\nu_r+(\beta_{1r}+\beta_{3r})\,r^2)+\cos{(4\theta)}\,\beta_{4r}\,r^2\,\nonumber\\
&&+\sin{(2\theta)} (\nu_i+(\beta_{3i}-\beta_{1i})\,r^2)+\sin{(4\theta)}\,\beta_{4i}\,r^2\large)\,,
\label{eqr}
\end{eqnarray}
\begin{eqnarray}
\dot{\theta}&=&\mu_i+\beta_{2i}\, r^2+\sin{(2\theta)}\,(-\nu_r+(\beta_{1r}-\beta_{3r})\,r^2)-sin{(4\theta)}\,\beta_{4r}\,r^2\nonumber\\
& &+\cos{(2\theta)}\,(\nu_i+(\beta_{1i}+\beta_{3i})\,r^2)+\cos{(4\theta)}\,\beta_{4i}\,r^2\,.
\label{eqtet}
\end{eqnarray}

The invariance ${\bf B} \rightarrow-{\bf B}$ amounts to the invariance $\theta\rightarrow\theta+\pi$. When the equation for $\theta$ does not have stationary solutions, the solution is oscillatory. Qualitative understanding is gained if we assume that the dynamics of $r$ can be adiabatically eliminated. In this limit, $\dot{r} \simeq 0$, we obtain from eq. \ref{eqr},
$r \simeq r_0 (\theta)$ such that we can write eq. \ref{eqtet} in the form
$\dot{\theta}=G_1(\theta)-G_2(\theta)$ and define 
\begin{eqnarray}
G_1(\theta)&=&\mu_i+\beta_{2i}\, r_0^2+\cos{(2\theta)}\,(\nu_i+(\beta_{1i}+\beta_{3i})\,r_0^2)\,,\nonumber\\
G_2(\theta)&=&\sin{(2\theta)}\,(\nu_r-(\beta_{1r}-\beta_{3r})\,r_0^2)+sin{(4\theta)}\,\beta_{4r}\,r_0^2-\cos{(4\theta)}\,\beta_{4i}\,r_0^2,
\label{defGi}
\end{eqnarray}
A fixed point  $\theta_c$ corresponds to an  intersection of the curves $G_1$ and $G_2$.  It is  stable if $G'_1(\theta_c)-G_2'(\theta_c)$ is negative, {\it i.e.} $G_1$ is larger than  $G_2$ for $\theta$ smaller than $\theta_c$. It is unstable otherwise.

When the solution is stationary, we obtain the relative importance of dipolar and quadrupolar components  from  $D=r_0\, \cos(\theta_c)$ and $Q=r_0\, \sin(\theta_c)$.  For positive $\mu_r$ and $\nu_r$, the equation has in general two pairs of solutions, one of which is stable and the other is unstable. It may have more solutions when non linearities associated to $\bar{A}^3$ are important (P\'etr\'elis \& Fauve 2008).

\begin{figure}
\begin{center}
\includegraphics[width=.5\textwidth]{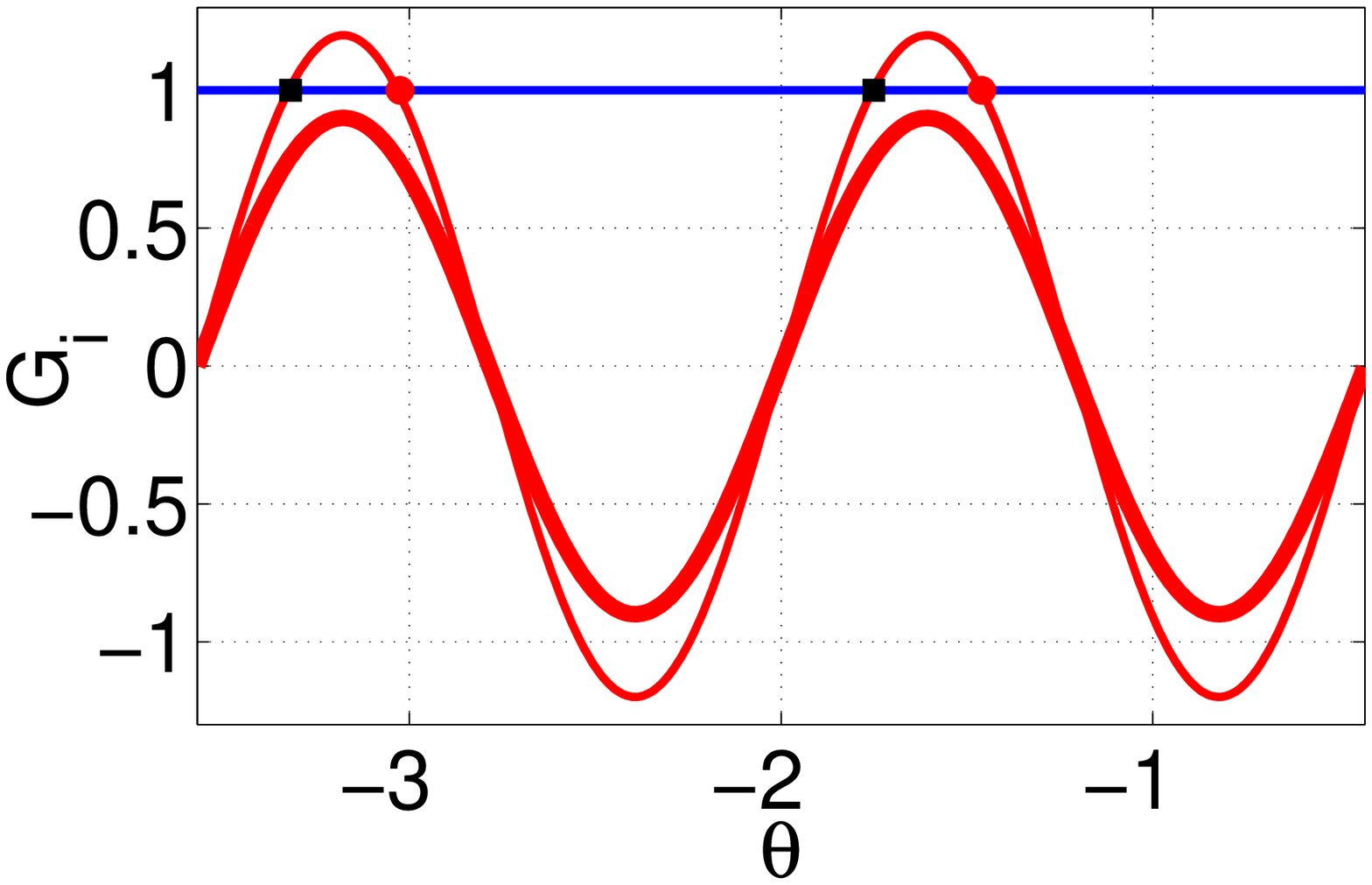}\includegraphics[width=.5\textwidth]{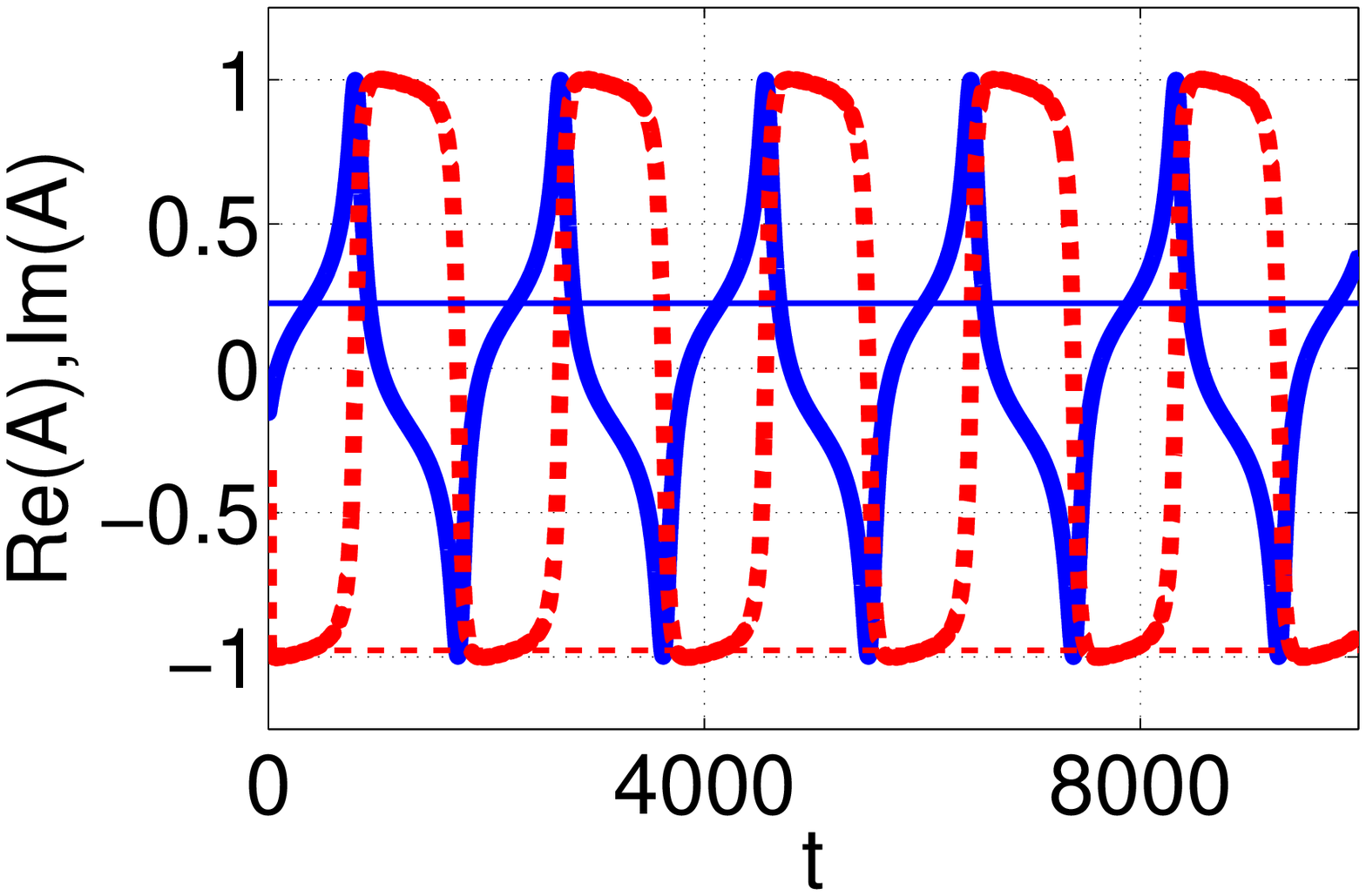}
\caption {Left: functions $G_i (\theta)$ that control the phase dynamics defined by eq. \ref{defGi}. We have chosen $\mu_r=1$. $G_1$ (blue line), $G_2$ (red line). The thin line corresponds to $\nu_r=1.2$.  Stable fixed points  ($\blacksquare$) and  unstable fixed points ($\bullet$). The thick line corresponds  to  $\nu_r=0.9$, there is no longer any fixed point and the solution is oscillatory. Right: time series of the solutions of eq. \ref{eqdipquad} for  $f=1.05$ (thick line),  $f=1.2$ (thin line);  dipolar component $Re(A)$ (continuous blue line), quadrupolar component $Im(A)$ (dashed red line).}
\label{figdem}
\end{center}
\end{figure}

When the values of the parameters are changed,  the positions of the stable and unstable fixed points evolve. If they collide, they disappear and the solution becomes oscillatory. As an example, we consider the case  $\mu_r=1$, $\beta_{2r}=-1$, $\mu_i=1$ and all the other coefficients are zero but $\nu_r$. The functions $G_1$ and $G_2$ are sketched in   fig. \ref{figdem}. For $\mu_i<\nu_r$, there are  $2\times2$  fixed points. If $\mu_i>\nu_r$, there are no fixed points  and the solution is oscillatory.

As said above, for the VKS experiment, we expect that $\nu_r$ changes  sign if the frequency difference $f$ increases from zero. When $\nu_r$ vanishes, if $\mu_i$ is larger than  $\nu_i$, the solution  is oscillatory. When $\nu_r$ reaches larger  negative values, a stationary solution can become stable again.  Therefore, we have a mechanism that leads to successive stationary, oscillatory and  again stationary states when f is increased. 

Assume that the terms that vary, $\mu_i$, $\nu_i$ and $\nu_r$, are small compared to $\mu_r$ which controls the amplitude of the fields. Then $r_0(\theta)$ weakly depends on $\theta$. In the phase space ($D=Re(A),  Q=Im(A)$),  the stationary solutions are points located on the limit cycle associated to the oscillatory solution.  

\begin{figure}
\begin{center}
\includegraphics[width=.6\textwidth]{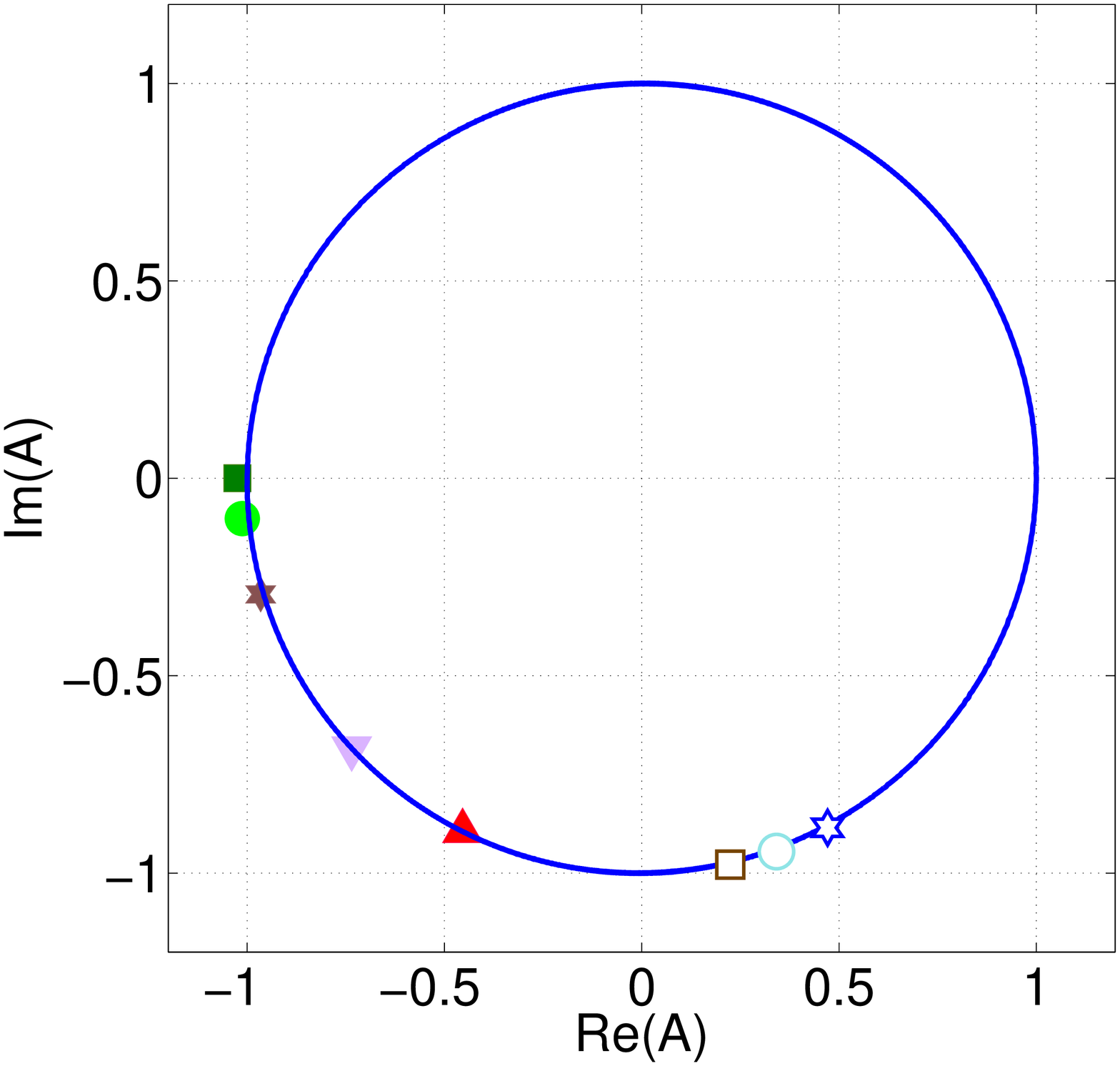}
\caption{Phase space representation  of the solutions of eq. \ref{eqdipquad}. The stationary solutions are represented by symbols located on the  limit cycle (obtained for $f=1$) and are associated to  increasing $\theta$ when $f$ increases from $0$ to $1.5$ ($f=0$, $0.5$, $0.75$, $0.9$, $0.95$, $1.2$, $1.4$, $1.475$).}
\label{fig9}
\end{center}
\end{figure}

To test this scenario, we numerically calculate the solutions of eq. \ref{eqdipquad} with the following coefficients: 
$\mu_r=1;\,\beta_{2}=-1$; $\beta_1=\beta_3=\beta_4=0$,
$\nu_r=0.05\,(1-1.29\,f^2+0.29\,f^4)$,
$\mu_i=0.0028\,(f+2\,f^3)$,
$\nu_i=0.0104\,(f-0.222\,f^3)$,
and $f$ is a parameter that varies between $0$ and $1.5$. Time series of the solution for various $f$ are presented in fig. \ref{figdem} (right). The phase space $(Re(A),\,Im(A))$ is shown in fig. \ref{fig9}.

When both modes are at the same distance from onset, $\nu_r\simeq 0$, the  solution is oscillatory if  $\mu_i^2>\nu_i^2+\nu_r^2$. Therefore, when $\nu_r$ changes sign, the system evolves from a stationary solution towards an oscillatory solution and then back  to a stationary solution. The limit cycle is generated by a saddle-node bifurcation and, slightly above onset, the slow phases of the cycle are very close to the former fixed points. This explains the form of the signal that displays a plateau or a slow evolution close to the former fixed points  (see fig. \ref{figdem} right).

Appropriate scaling of the terms $\mu_i$, $\nu_i$ and $\nu_r$ as functions of $f$ is required so that, the system is close to the quadrupolar solution before the bifurcation to the limit cycle takes place, the stationary solutions are points located on the limit cycle associated to the oscillatory solution, and when a stationary solution is recovered after the oscillatory regime, its phase $\theta_c$ increases for a large range of $f$. When these requirements are fulfilled, our model exactly reproduces the behaviour of the first bifurcations displayed by the VKS experiment when $F_1-F_2$ is increased from zero (see figure 3 in Ravelet \textit{et al.} (2008).

\section{Scenario of the bifurcation and mechanism for reversals}

\subsection{A generic bifurcation}

In the VKS experiment, most dynamical regimes are reached through bifurcations from a stationary solution $B_s$. At threshold, the generated limit cycle connects $B_s$ to  $-B_s$. It can be time-periodic, as it is the case for the regime described above, or random, and then looks like reversals of the Earth magnetic field. 
 
This kind of bifurcation is not classical (Arnold 1982). Actually, it  is enforced by the symmetries of the problem and we can prove the following result. Consider a planar system invariant under  the transformation ${\bf B}\rightarrow-{\bf B}$ and with two different and non zero stationary solutions. One of the fixed points is unstable, ${\bf B}_u$ and the other one is stable, ${\bf B}_s$. The collision between the two fixed points generates a cycle that connects the collision point with its opposite, see fig. \ref{figmodrenv}.   
This result can be understood as follows: the solution $B=0$  is unstable with respect to  the two different fixed points, and their opposite. It is an unstable point, whereas one of the two bifurcating solutions is a stable point, a node,  and the other is a saddle. If the saddle and the node collide, say at $B_c$, what happens to initial conditions located close to these points? They cannot be attracted by $B=0$ which is unstable and they cannot reach other fixed points since they just disappeared. Therefore the trajectories describe 
a cycle.  The associated orbit contains $B=0$ since, for a planar problem, in any orbit, there is a fixed point. Suppose that the orbit created from $B_c$ is different from the one created by $-B_c$. These orbits being images by the transformation ${\bf B}\rightarrow-{\bf B}$, they must intersect  at some point. Of course, this is not possible for a planar system because it would violate the unicity of the solutions. Therefore, there is only one cycle that connects points close to $B_c$ and $-B_c$.

\subsection{Regimes in the neighborhood of this bifurcation}

Two regimes can result from that scenario. One of the regimes is located above the onset of the saddle-node bifurcation and we now discuss its properties in the absence of fluctuations. The period $T$ of the limit cycle diverges when the distance to the bifurcation threshold, $\epsilon$, vanishes. 
Indeed, for $\epsilon$ very small, the period $T$ is controlled by the time spent close to the former fixed points. There, the dynamics can be written, up to a change of variables, $\theta = \theta_c + \phi$ ($\phi \ll 1$),
\begin{equation}
\dot{\phi}=\epsilon+C \phi^2\,,
\label{sadnod}
\end{equation}
where $C$ is a constant. This equation is the normal form of a saddle-node bifurcation (Arnold 1982). 
Integration of eq. \ref{sadnod} from the initial condition $\phi(t=0)=0$ gives $\phi=\sqrt{\epsilon/C}\,\tan{(\sqrt{\epsilon C} t)}$ from which the relation $T\simeq 2 \pi/\sqrt{\epsilon C}$ is easily derived.\\

\begin{figure}
\begin{center}
\includegraphics[width=.60\textwidth]{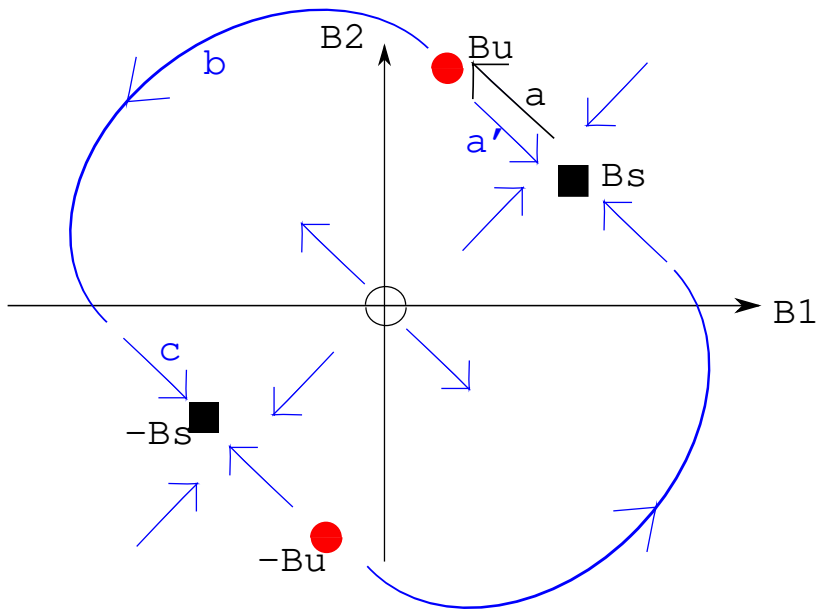}
\caption{Phase space for reversals and excursions. The stable fixed  points $B_s$ are black squares. The unstable fixed points $B_u$ are red circles. Initially (phase a), fluctuations drive the system away from the stable point (fluctuations driven dynamics sketched in black). Then, two situations can take place: in the first one, the system returns to $B_s$ (phase a', deterministic dynamics sketched in blue). In the second one, the system evolves to $-Bs$ (phases b and c, deterministic dynamics in blue) and the system undergoes a reversal with an overshoot for the component $B_1$ during phase c. }
\label{figmodrenv}
\end{center}
\end{figure}

The second regime is located below the bifurcation threshold and requires the presence of fluctuations.   
The system is attracted by one of the fixed points that is locally stable, $\pm B_s$. However, $B_s$ is very close to an unstable fixed point, $B_u$ (see fig. \ref{figmodrenv}). Fluctuations can push the system to the unstable fixed point and once that point is reached, the system is attracted by the stable fixed point of opposite sign $-B_s$. It remains close to  $-B_s$ until fluctuations trigger a new evolution toward $B_s$. This is a mechanism of excitability, that results in reversal-like behavior when the system is invariant with respect to  ${\bf B}\rightarrow-{\bf B}$. If we approximate again the dynamics by the evolution of the phase $\phi$, we can write, 
\begin{equation}
\dot{\phi}=\epsilon+C \phi^2+\zeta(t) \,,
\label{sadnodb}
\end{equation}
where $\zeta$ is a random noise of intensity $D$ that describes the effect of the fluctuations. This equation is valid close to the bifurcation threshold for $\phi \ll 1$ and we consider here $\epsilon\le 0$. It can be written as the evolution equation of an overdamped motion in a metastable potential well $V(\phi)=-\epsilon \phi - \phi^3/3$.
Between reversals, the system fluctuates in the vicinity of the stable fixed point. The durations of these ``fixed polarity phases" are random. Classical results on the exit time from a metastable state show that they are exponentially distributed with a characteristic time that depends on the fluctuation intensity and the distance to bifurcation. Thus,  $P(T)\propto\exp{(-T/<T>)}$, with $<T>\propto \exp{(\Delta V/D)}$ when the noise intensity $D$ is small compared to the difference of energy between the fixed points $\Delta V= V(\phi_{max})-V(\phi_{min})=4 C^{-1/2}(-\epsilon)^{3/2}/3$.  This result is valid close to the saddle-node threshold so that eq. \ref{sadnodb} is pertinent but not too close so that the fluctuations do not drive too frequent escapes. Nevertheless, this is the appropriate regime if we want to describe reversals that occur on time scales very long compared to the characteristic time scales of the deterministic dynamics.

\subsection{A model}

\begin{figure}
\begin{center}
\includegraphics[width=.5\textwidth]{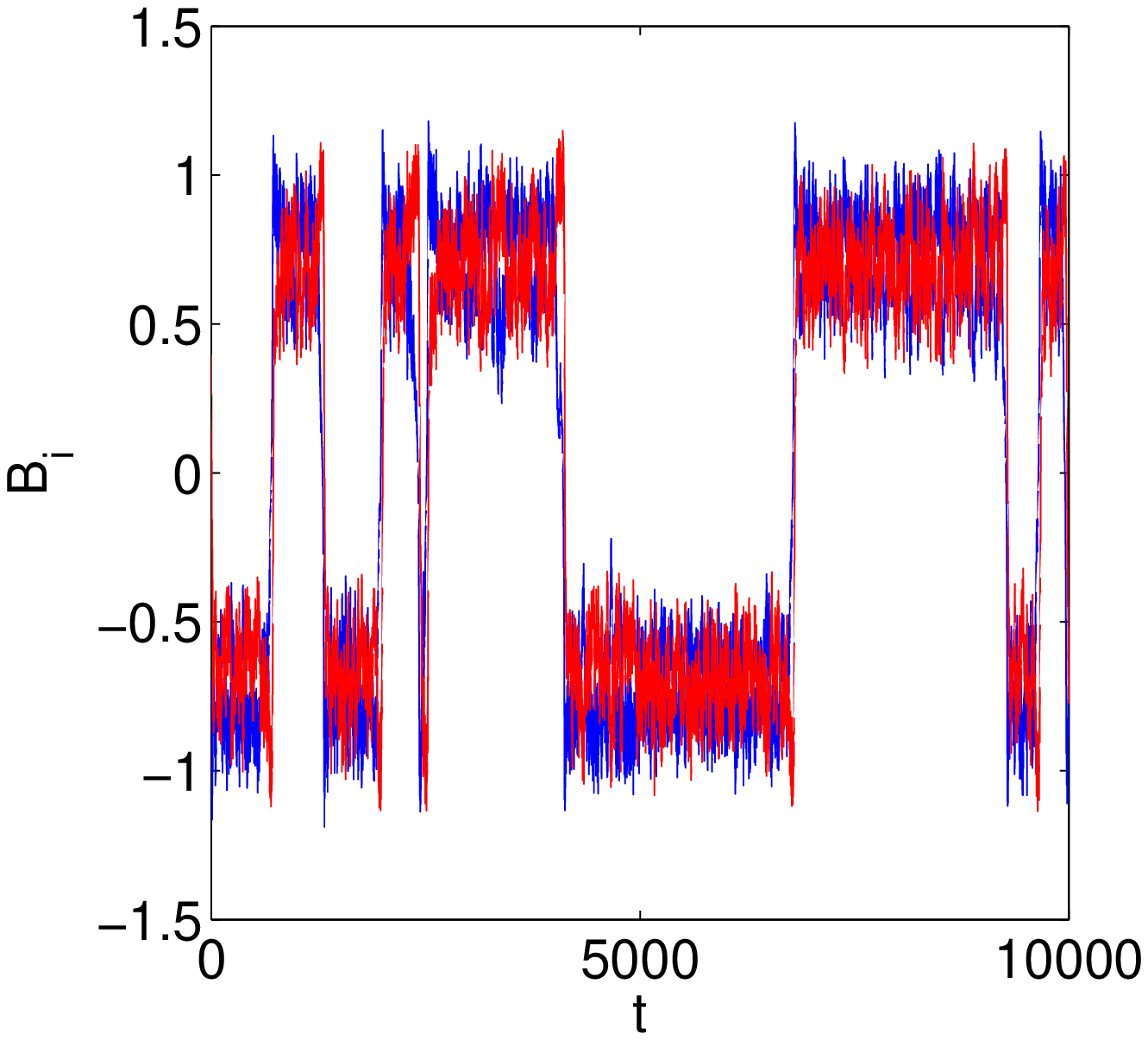}\includegraphics[width=.5\textwidth]{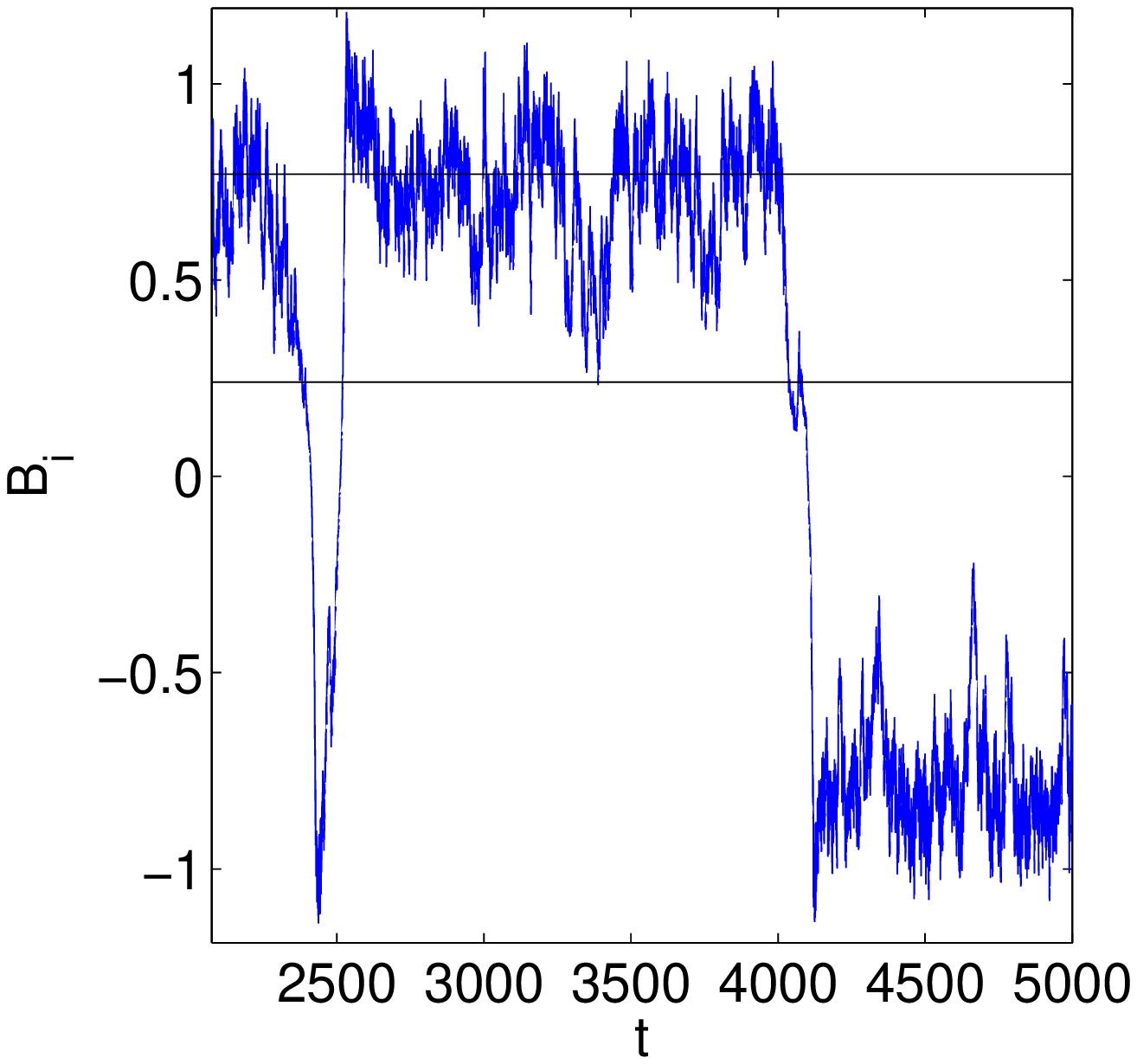}
\caption{Left: time series of the  solution of eq. \ref{eqnc} for $br_1=br_4=0.1$. Right: zoom on the first component. The horizontal lines represent the stable solution (upper line) and the  unstable solution (lower line) in the absence of noise.}
\label{figrenv}
\end{center}
\end{figure}

We use now a simple model in order to describe the effect of turbulent fluctuations on the dynamics of the two magnetic modes governed by \ref{eqdipquad}. We change notation, $A = B_1 + i B_2$ to stress that the two modes do not need to be a dipole and a quadrupole in general. 
\begin{eqnarray}
\dot{B_1}&=&(\mu_r+\nu_r) B_1+ (\nu_i-\mu_i) B_2\nonumber\\
& &+C_{11}\,B_1^3+C_{21}\,B_1^2\,B_2+C_{31}\,B_1\,B_2^2+C_{41}\,B_2^3+br_1 \zeta_1(t) B_1+br_2 \zeta_2(t) B_2\,,\nonumber\\
\dot{B_2}&=&(\mu_r-\nu_r) B_2+ (\nu_i+\mu_i) B_1\nonumber\\
& &+C_{12}\,B_1^3+C_{22}\,B_1^2\,B_2+C_{32}\,B_1\,B_2^2+C_{42}\,B_2^3+br_3 \zeta_3(t) B_1+br_4 \zeta_4(t) B_2\,.
\label{eqnc}
\end{eqnarray}
The nonlinear coefficients $C_{i\,j}$ are derived from those of eq. \ref{eqdipquad}. Turbulent fluctuations are modelled by the terms  $\zeta_i$ that are independent gaussian white noises (with the Stratanovich interpretation). We take $\mu_r=1$, $\nu_r=0.05$, $\mu_i=0.045$, $\nu_i=0.025$ and $\beta_{2r}=-1$. The other non-linear coefficients are set to zero. For simplicity, we set $br_2=br_3=0$ and take $br_1 = br_4 =0.1$.  Time series of the solutions are displayed in fig. \ref{figrenv}.

We observe that a reversal consists of two phases. In the first phase, the system evolves from the stable point $B_s$ to the unstable point $B_u$. The deterministic part of the dynamics acts against this evolution and the fluctuations are the motor of  the dynamics. That phase is thus slow. In the second phase, the system evolves from $B_u$ to $-B_s$, the deterministic part of the dynamics drives the system and this phase is faster. Both phases can be identified in fig. \ref{figrenv} (right).

The behaviour of the system close to $B_s$ depends on the local flow. Close to the saddle-node bifurcation,  the position of $B_s$ and $B_u$  defines the slow direction of the dynamics. If a component of $B_u$ is smaller than the corresponding one of  $B_s$, that component displays an overshoot at the end of a reversal. In the opposite case, that component will increase at the beginning of a reversal. For instance, in the phase space sketched  in figure \ref{figmodrenv}, the component $B_1$ decreases at the end of a reversal and the signal displays an overshoot.  The component $B_2$ increases just before a reversal.

For some fluctuations, the second phase does not connect $B_u$ to $-B_s$ but to $B_s$. It is an aborted reversal or an excursion in the context of the Earth dynamo.  Note that during the initial phase, a reversal and an excursion are identical. In the second phase, the approaches to the stationary phase differ because the trajectory that links $B_u$ and $B_s$ is different form the trajectory that links $B_u$ and $-B_s$. 
In particular, if the reversals display an overshoot this will not be the case of the excursion (see fig. \ref{figrenv} (right)  and the sketch in fig. \ref{figmodrenv}).

\section{Conclusion}

We have studied dynamical regimes that can arise when two axisymmetric magnetic eigenmodes are coupled. Symmetry considerations allow to identify properties of the magnetic modes and, in some cases, put constraints on the coupling between the modes.  We have shown that when a discrete symmetry is broken by the flow that generates the magnetic field, the coupling between an odd and an even magnetic mode (with respect to the symmetry) can generate a bifurcation from a stationary state to a periodic state. This behaviour is generic when a saddle-node bifurcation occurs in a system that is invariant under ${\bf B} \rightarrow-{\bf B}$. Close to the the bifurcation threshold, fluctuations drive the system into a state of random reversals that connect a solution $B_s$ to its opposite $-B_s$.  
This scenario provides a simple explanation for many features of the dynamics of the magnetic field observed in the VKS experiment: alternation of stationary and time dependent regimes when a control parameter is varied, continuous transition from random reversals to time periodic ones, characteristic shapes of the time recordings of reversals versus excursions. 

Although the discrete symmetry involved for the flow in the Earth core is different from the one of the VKS experiment, a similar analysis can be performed for the geodynamo (P\'etr\'elis  \textit{et al.} 2008). 

More generally, our scenario can be applied to purely hydrodynamic systems. Cellular flows driven by thermal convection (Krishnamurti \& Howard 1981) or by volumic forces (Sommeria 1986) display a transition for which a large scale circulation is generated on a smaller scale turbulent background. This large scale flow can display random reversals, very similar to the ones observed for the magnetic field. A model analogue to the present one, can explain how this large scale field can reverse without the need of a very energetic turbulent fluctuation acting coherently in the whole flow volume.

\end{document}